# Calibration of the Gamma-RAy Polarimeter Experiment (GRAPE) at a Polarized Hard X-Ray Beam

P. F. Bloser[1], J. S. Legere, M. L. McConnell, J. R. Macri, C. M. Bancroft, T. P. Connor, J. M. Ryan

Space Science Center, University of New Hampshire, Durham, NH 03824, USA

**Abstract**

The Gamma-RAy Polarimeter Experiment (GRAPE) is a concept for an astronomical hard X-ray Compton polarimeter operating in the 50 – 500 keV energy band. The instrument has been optimized for wide-field polarization measurements of transient outbursts from energetic astrophysical objects such as gamma-ray bursts and solar flares. The GRAPE instrument is composed of identical modules, each of which consists of an array of scintillator elements read out by a multi-anode photomultiplier tube (MAPMT). Incident photons Compton scatter in plastic scintillator elements and are subsequently absorbed in inorganic scintillator elements; a net polarization signal is revealed by a characteristic asymmetry in the azimuthal scattering angles. We have constructed a prototype GRAPE module containing a single CsI(Na) calorimeter element, at the center of the MAPMT, surrounded by 60 plastic elements. The prototype has been combined with custom readout electronics and software to create a complete "engineering model" of the GRAPE instrument. This engineering model has been calibrated using a nearly 100% polarized hard X-ray beam at the Advanced Photon Source at Argonne National Laboratory. We find modulation factors of 0.46 ± 0.06 and 0.48 ± 0.03 at 69.5 keV and 129.5 keV, respectively, in good agreement with Monte Carlo simulations. In this paper we present details of the beam test, data analysis, and simulations, and discuss the implications of our results for the further development of the GRAPE concept.

[1] Corresponding author:
Address: University of New Hampshire, 8 College Road, Durham, NH 03824, USA
Phone: +1 603 862 0289
Fax: +1 603 862 3584
Email: Peter.Bloser@unh.edu

## 1. Introduction

An important but so far under-exploited tool in investigating high-energy astrophysical phenomena is polarimetry. Many emission processes that can generate X-ray and gamma-ray photons, including synchrotron radiation, Compton scattering, and electron-proton bremsstrahlung, can also result in the linear polarization of those photons (e.g., [1]). The level of polarization, however, depends on the precise emission geometry. In addition, the energy dependence of the polarization can provide clues to the emission mechanisms that may be operating. Polarization measurements therefore have the potential to tell us much about both the physical mechanisms and source geometries responsible for the observed high-energy emissions, especially when combined with the more traditional astronomical techniques of imaging, spectroscopy, and timing.

Two examples of high-energy astrophysical phenomena for which high-energy polarimetry promises to yield new insights are gamma-ray bursts (GRBs) and solar flares. GRBs are brief, intense flashes of gamma rays randomly distributed across the sky that are widely believed to herald the formation of a black hole, whether via supernova explosion or the merger of a compact binary [2, 3]. In either case, the prompt gamma-ray emission is believed to result from particles accelerated by shocks in a highly relativistic jet aimed at the observer. Despite extensive observational efforts (e.g., CGRO, HETE-2, BeppoSAX, INTEGRAL, Swift, and others), several key properties of GRBs remain poorly understood and are difficult or even impossible to infer with the spectral and timing information currently collected. Polarimetry will offer a fresh perspective on the nature of the GRB "central engine." For example, it is speculated that strong magnetic fields play an essential role in launching the relativistic jet. It is not clear, however, whether these magnetic fields are globally ordered or tangled and chaotic. Waxman [4] describes how each scenario could lead to polarized emission: 1) In the "physical" scenario, a globally ordered magnetic field permeating the emission region could lead to high levels of polarization via synchrotron radiation. The maximum polarization, ~70%, would be seen along the jet axis, but levels greater than 20% would be seen for most viewing angles within the jet opening angle. 2) Alternatively, in a "geometric" scenario, a jet with a tangled magnetic field could produce a high polarization fraction only if it were viewed very close to the edge of the jet, owing to relativistic beaming effects. Therefore, given a large sample of GRBs with a random distribution of viewing angles, a high fraction of events with significant levels of polarization would support the physical scenario over the geometric scenario. We also note that any energy dependence of the polarization fraction could be a clue to the GRB radiation mechanism, since for inverse Compton scattering the degree of polarization varies with energy, whereas for synchrotron emission the polarization degree is constant [1].

Solar flares represent a process of explosive energy release in the magnetized plasma of the solar corona. Flares accelerate ions up to tens of GeV and electrons to hundreds of MeV, releasing as much as $10^{33}$ ergs in the process (see, e.g., [5]). The accelerated 10 – 100 keV electrons appear to contain a significant fraction, perhaps the bulk, of this energy. The details of how the Sun releases this energy, presumably stored in the magnetic fields of the corona, and how it rapidly accelerates electrons with such high efficiency, and to such high energies, is presently unknown. Polarization measurements are expected to be useful in determining the beaming, or directivity, of solar flare electrons, a quantity that may provide important clues about electron acceleration and transport. Measurements at energies above 50 – 100 keV are expected to be most useful [6], where large linear polarizations of the electron bremsstrahlung radiation are expected due to the anisotropy of the electrons (e.g., [7, 8]) and contamination from thermal (non-polarized) emission is avoided.

The use of polarimetry in high-energy astronomy has, until recently, been limited to energies below ~20 keV [e.g., 9 – 11]. However, the hard X-ray/soft gamma-ray energy band (roughly 50 keV – 1 MeV) is better suited for studying the non-thermal processes at work in GRBs and solar flares. In this range, polarization measurements have finally been reported in recent years for GRBs [12 – 15] and solar flares [16 – 20]. Unfortunately, these hard X-ray measurements have generally been of limited statistical significance, in some cases generating significant controversy (e.g., [21, 22]). This is largely due to the fact that most of these reported measurements were made using instruments that were not designed specifically for high-energy polarimetry.

Significant progress in the field of high-energy polarimetry will require instruments specially designed for the task. To that end, we have, over the past ten years, developed a concept for a dedicated hard X-ray (50 – 500 keV) polarimeter currently known as the Gamma-RAy Polarimeter Experiment (GRAPE) [23 – 30]. We note that several other hard X-ray polarimeter designs have been discussed in the literature in recent years as well (e.g., [31 – 33]), which underscores the scientific interest in these measurements. GRAPE is optimized for the study of GRBs, solar flares, and other energetic transients over a wide field of view from either a satellite or long-duration balloon platform. For bright events, the polarization angle and degree can be measured as a function of both time and energy, allowing tight constraints to be placed on physical models. A typical implementation of the GRAPE concept employs an array of identical modules, each of which operates as an independent Compton polarimeter. More sophisticated applications are possible as well: GRAPE could easily be adapted into a high-energy imaging system using, for example, rotation modulation collimators. A prototype GRAPE module has been tested in the laboratory, calibrated at a polarized X-ray beam at Argonne National Laboratory, and flown on an engineering balloon test flight from Palestine, TX. In this paper we describe in detail the beam calibration and data analysis, and compare the measured results with Monte Carlo simulations. The GRAPE balloon flight will be the subject of a future paper.

## 2. The GRAPE Instrument

### *2.1 Principle of Operation*

The design of the GRAPE polarimeter module is based on the fact that, in Compton scattering, photons are preferentially scattered at right angles to the incident electric field vector (the polarization vector). If the incident photon beam carries a net linear polarization, the distribution of azimuthal scattering angles will be asymmetric. Specifically, if the polarization vector lies along azimuthal angle $\phi_0$ then the distribution of azimuthal scatter angles $\phi$ is given by [1]

$$C(\phi) = A\cos\left[2\left(\phi - \phi_0 + \frac{\pi}{2}\right)\right] + B, \tag{1}$$

where $A$ and $B$ are constants. The degree of asymmetry, which is a direct measure of the polarization fraction, is given by the modulation factor $\mu$, which is defined as

$$\mu = \frac{C_{\max} - C_{\min}}{C_{\max} + C_{\min}} = \frac{A}{B}, \tag{2}$$

where $C_{max}$ and $C_{min}$ refer to the maximum and minimum counts, respectively, in the azimuthal modulation profile. The modulation factor decreases with increasing photon energy, and is largest for Compton scattering angles (i.e., polar angles) near 90º. In order to measure the degree of polarization, it is necessary to know the modulation factor for 100% polarized radiation, $\mu_{100}$; for a measured $\mu$ the polarization fraction Π is then simply given by

$$\Pi = \frac{\mu}{\mu_{100}}. \tag{3}$$

For high-sensitivity polarization measurements it is desirable for $\mu_{100}$ to be as large as possible.

In general, a Compton polarimeter consists of two detectors, one that measures the position of the Compton scatter and the energy of the recoil electron, and one that measures the direction and energy of the scattered photon. The first, the scattering detector, should be of a low-density, low-Z material to maximize the probability of Compton scattering relative to absorption. The second, the calorimeter, should be of a denser, higher-Z material in order to efficiently absorb the full energy of the scattered photon. The position resolution of these detectors determines the accuracy to which the scatter angle $\phi$ can be measured. In practice, the finite spatial resolution of the detectors degrades the value of $\mu_{100}$ for a given polarimeter and limits the polarization sensitivity. Similarly, the energy resolution of the detectors limits the accuracy with which energy-dependent polarization can be measured. Finally, the geometric layout of the scattering and calorimeter detectors can introduce artificial, systematic asymmetries into the measured azimuthal scattering pattern that can either mimic or mask a true polarization signal. To remove these effects, one of two things is necessary: 1) to know, with high accuracy, the azimuthal response of the polarimeter to completely unpolarized radiation, so that subsequent measurements can be corrected; or 2) to physically rotate the polarimeter during measurements, so that geometric asymmetries are averaged out.

Based on these considerations, the design of the basic GRAPE polarimeter module is shown schematically in Fig. 1. The module consists of an 8 × 8 array of optically independent, 5 mm × 5 mm × 50 mm scintillator elements read out by a multi-anode photomultiplier tube (MAPMT). Two types of scintillator are used. For the scattering detector elements, located in the central 6 × 6 sub-array, low-Z plastic scintillator provides an effective medium for Compton scattering. The plastic elements are surrounded by 28 calorimeter elements made of high-Z inorganic scintillator. Each detector element is optically coupled to one of the 64 anodes of a Hamamatsu H8500 MAPMT. Valid polarimeter events are those in which an incident photon Compton scatters in one plastic element and is subsequently absorbed in one calorimeter element; such events are identified as a coincident trigger in exactly one plastic anode and one calorimeter anode. The azimuthal scatter angle $\phi$ is determined by the relative locations of the centers of the hit scintillator elements (Fig. 1, left). A polarization signal is found by creating a histogram of the measured $\phi$ values and, after correcting for geometric effects, fitting it with Eq. 1 (Fig. 1, right). The modulation factor is found from the fit parameters *A* and *B* (Eq. 2), which, assuming $\mu_{100}$ is known for the relevant energy band, permits calculation of the polarization fraction Π via Eq. 3. The polarization angle is given by the fitted value of $\phi_0$ (Fig. 1, right; Eq. 1). The compact, square design allows many modules to be packed tightly together into an array, making for a modular, flexible instrument.

### *2.2 Prototype Detector Module*

We have developed and tested a series of prototype GRAPE modules [23 – 30]. The latest [29, 30], referred to here as the third Science Model (SM3), uses, for reasons of cost, a slightly different scintillator layout (Fig. 2) than that described in Sec. 2.1. A single, 10 mm × 10 mm × 50 mm calorimeter element, made of CsI(Na) (manufactured by Proteus, Inc.), is located at the center of the module. The calorimeter is surrounded by 60 scattering elements (5 mm × 5 mm × 50 mm) made of EJ-204 plastic (Eljen Technology). Each scintillator element is wrapped in VM2000™ reflective material (3M) and optically cemented to the entrance window of a H8500 MAPMT (Hamamatsu Corp.). The H8500 provides an 8 × 8 array of independent, 5-mm anodes on a 6-mm pitch. Each plastic element corresponds to a single anode, whereas the calorimeter covers four anodes. The data acquisition system for SM3 is based on laboratory VME and NIM electronics.

The choice of CsI(Na) for the SM3 calorimeter was motivated by this scintillator's good spectral match to traditional bialkali photocathodes, good light output, and relatively low cost. The energy resolution (roughly 7.3% FWHM at 662 keV) is sufficient for the study of continuum sources such as GRBs and solar flares. An additional advantage is that the CsI(Na) decay time, ~600 ns, is significantly different from that of the EJ-204 plastic (1.8 ns). In early testing it was found that optical cross talk between calorimeter and plastic elements was a significant problem in SM3 [29]. For the Engineering Model, however, we have been able to circumvent this issue by making use of the differing CsI(Na) and plastic decay times to reject cross talk events using pulse shape discrimination (Sec. 2.3).

The SM3 module, like the earlier prototypes, has been tested in the laboratory using a partially polarized hard X-ray beam created by scattering 662 keV photons from a $^{137}$Cs source at 90º in a block of plastic scintillator [27 – 29]. This creates a beam with a polarization fraction of ~55 – 60% at an energy of ~288 keV. Data were collected with SM3 at various rotation angles relative to the polarization angle. The polarized data were corrected for geometric effects (see Sec. 2.1) using an unpolarized 356 keV beam from a $^{133}$Ba source. The measured value of $\mu$ was 0.35 ± 0.04, and the derived beam polarization was 56 ± 9% [29] using a value of $\mu_{100}$ calculated from Monte Carlo simulations (see Sec. 5). For this measurement the twelve plastic elements closest to the calorimeter were excluded due to optical cross talk [29].

### *2.3 Engineering Model*

In order to further develop the GRAPE polarimeter design beyond the laboratory prototype stage and prepare for balloon-flight applications, an Engineering Model (EM1) was developed (Fig. 3). EM1 combines the SM3 prototype detector with custom readout electronics and data acquisition software running on a dedicated PC/104 computer. The electronics boards fit within the footprint of the MAPMT, allowing for multiple polarimeter modules to be tiled into a compact array for a future balloon-borne science instrument. The design was intentionally developed without the use of custom ASICs to assure flexibility with evolving balloon instrument configurations while minimizing the cost of future modifications.

The data acquisition (DAQ) electronics and software are shown schematically in Fig. 4. The readout is controlled by a PIC18F4620 microprocessor, which sets the readout mode ("polarimeter mode," "singles mode," or "rates mode"), collects the data from each valid event, and sends the data to a PC/104 computer via an RS-232 interface. "Polarimeter mode" requires a coincidence between the calorimeter and a user-selectable number of plastic elements (1 – 3 for the beam calibration). "Singles mode" records a stream of pulse heights for one selected channel and is used for calibration. "Rates mode" simply reports the count rate in all channels. The PC/104 runs custom DAQ software under Linux that communicates with the PIC processor via an RS-232 card, commands the module to run in a given mode for a given amount of time, and writes formatted data files to the hard disk containing, for each event, a timestamp, the number of elements hit, and the pulse height for the calorimeter and each plastic element. The system can be run remotely via an Ethernet connection to the PC/104.

The 64 anode outputs of the MAPMT are fed into four main readout boards with 16 channels each (shown in the center of Fig. 3). The signals first pass through an "analog channel" consisting of a charge pre-amplifier, a constant-fraction discriminator (CFD) for the calorimeter or a pulse-shape discriminator for the plastic elements, a Gaussian shaping filter, and a sample-and-hold. The discriminator threshold of each channel is individually set by the PIC processor. Each channel that exceeds threshold activates the corresponding channel "Event_N" line. The number of plastic elements above threshold ("#_Events") is encoded through a simple analog sum of the channel event lines and is one of the factors used in recognizing a valid event; for the beam test described below, we accepted 1, 2, or 3 plastic hits. The outputs of the calorimeter CFD and plastic discriminators on all four boards are combined ("CAL_Event" and "Plastic_Event," respectively) and fed into a timing and control circuit on the module processor board (visible at the bottom of Fig. 3). An event in any plastic or calorimeter line triggers the event line ("Event") to the processor.

The relative timing of the signals is used to implement an effective "pulse-shape discrimination" (PSD) in order to eliminate optical cross talk. The "Plastic_Event" signal is delayed by an amount that depends on the pulse shape. A true plastic pulse, with a decay time of ~1.8 ns, is detected substantially faster than a CsI(Na) pulse, with a decay time of ~600 ns. An additional delay of ~600 ns, corresponding to the rise time of the calorimeter, is then imposed. The delayed plastic and calorimeter signals are fed into a coincidence circuit with a window of ~100 ns. Any signal from a plastic anode caused by cross talk from the CsI(Na) will be delayed beyond the coincidence window.

A coincidence between the CsI(Na) and the correct number of plastic elements indicates a valid event. A "gate" signal is first generated to prevent prior and subsequent interactions in the detectors from triggering the electronics. The PIC processor then reads out both the "Event" and "Peak" multiplexers in order to record both the channel numbers and held pulse heights, respectively, of all hit channels. The pulse heights of the four calorimeter anodes are summed. The event data is sent to the PC/104 via the RS-232 interface, and a reset pulse is then sent to ready the system for the next trigger. The RS-232 interface is currently the limiting factor on the speed of the readout; the maximum valid event rate in polarimeter mode is ~100 cts $s^{-1}$. (For future science instruments the RS-232 interface will be replaced with a high-speed LVDS serial link that, in combination with data buffering, will increase the readout rate by nearly a factor of ten.) For invalid triggers the system is reset in less than 5 μs.

The EM1 detectors and electronics were calibrated in the laboratory using radioactive isotopes. Each plastic element was calibrated in singles mode using two collimated gamma-ray sources, $^{109}$Cd (22 keV) and $^{241}$Am (60 keV), in order to determine the gain and set the energy threshold. For most plastic channels the threshold was set at 6 keV; about ten channels, however, had a higher level of low-energy noise and required that the threshold be set at 10 – 12 keV. The CsI(Na) calorimeter was calibrated in singles mode using the same two sources plus $^{57}$Co (122 keV), and its threshold was set at 30 keV. The dynamic range of the EM1 electronics imposed an upper limit of ~300 keV on the total energy measurement; for future implementations this will be extended to at least 500 keV.

In addition to the beam test described in the following sections, EM1 was flown by NASA's Columbia Scientific Balloon Facility on an engineering balloon test flight from Palestine, TX, in June 2007. The detector and DAQ system, in combination with active charged particle shields and additional housekeeping hardware, performed very well throughout the flight, demonstrating that the GRAPE concept is mature and appropriate for long-duration balloon or space-based applications. The details and results of the balloon flight will be the subject of a forthcoming paper. Here we concentrate on the beam test results.

## 3. Beam Calibration at APS

The GRAPE EM1 Engineering Model was calibrated using a hard X-ray beam at the Advanced Photon Source (APS) at Argonne National Laboratory from December 15 – 19, 2006. The goals of the beam campaign were: 1) to measure the modulation factor of the GRAPE instrument for nearly 100% polarized radiation at multiple energies; 2) to validate our Monte Carlo simulation framework; and 3) to gain experience with the transportation and operation of the GRAPE hardware in preparation for the balloon test flight.

The GRAPE EM1 was installed in the Midwest Universities Collaborative Access Team (MUCAT) 6ID-D beam line at APS. The experimental setup is shown schematically in Fig. 5. The instrument was exposed to linearly polarized X-ray beams at two energies, 69.5 keV and 129.5 keV. The polarization vector was horizontal, and the polarization fraction was calculated to be 97 ± 2%. The energy spread in the beam was less than 0.1% at both energies. It was necessary to greatly reduce the beam flux in order not to overwhelm the EM1 electronics. This was accomplished by de-tuning one stage of the double-scattering silicon (331) monochromator, employing a tungsten collimator, and including a series of up to seven 3-mm iron attenuators, and additional copper and tantalum attenuators as necessary. The resulting beam was less than 1 mm in diameter. It was therefore necessary to expose individual plastic scattering elements to the beam one at a time by means of an X-Y translation table (provided by APS). The instrument was positioned roughly using a laser, and the alignment was fine-tuned by monitoring the count rate in individual plastic elements as they were scanned through the beam. The target coincidence count rate in polarimeter mode was ~10 cts $s^{-1}$ in order to avoid significant dead time (see Sec. 2.3). Since the efficiency for coincident events varies across the face of the polarimeter module, being much lower for plastic elements near the edge of the instrument, the number of attenuators was adjusted as needed, resulting in greatly differing input intensities for each position. In addition, the overall beam intensity declined by roughly 30% over the course of twelve hours due to gradual losses in the main electron storage ring; the ring was re-filled twice daily. The incident intensity on each plastic element therefore differed greatly, and so was monitored by means of a 1 cm-thick plastic scintillator paddle. Spectra from the beam monitor were recorded by an Amptek MCA8000A multi-channel analyzer (MCA). We note that the beam monitor was only intended to measure the relative intensity of the beam for different plastic elements at a given energy, not the absolute intensity. The beam monitor gain and threshold were adjusted for the two input energies, as was the attenuator configuration, and so we do not attempt to calculate either the absolute or relative efficiency of GRAPE as a function of energy.

Due to time constraints, only half of the 60 plastic elements were exposed to the beam. The plastic elements were scanned in a checkerboard pattern (Fig. 6), and roughly 10,000 polarimeter-mode events (calorimeter together with 1 – 3 plastics) were recorded on the PC/104 at each position. A beam monitor file was recorded on the MCA for each position as well. The total exposure time of each file was noted. After completing a scan EM1 was rotated 90º (Fig. 6), and the scan was repeated for the same plastic elements in the new orientation. In this way we were able to create an "unpolarized" exposure by combining the data from both orientations relative to the polarization vector; this unpolarized data was used for the removal of systematic geometric effects (see Sec. 4). This procedure was repeated for the 69.5 keV and 129.5 keV beam runs.

Upon our return from the beam campaign, two additional scans were made in the laboratory using collimated radioactive isotopes. The same plastic elements illuminated at APS were exposed to $^{241}$Am (60 keV) and $^{57}$Co (122 keV) sources under identical operating conditions in order to obtain additional unpolarized data sets for correcting the 69.5 keV and 129.5 keV polarized beam runs, respectively. Due to time constraints the same number of events, ~10,000, were recorded at each position as for the polarized runs, although in principle it is desirable to have significantly better statistics for the unpolarized data.

## 4. Data Analysis and Results

In order to create and analyze azimuthal scatter angle histograms for the APS beam runs, it was necessary to properly combine the data, accounting for the varying beam intensity. For each exposed scan position, we created two histograms, one for azimuthal scatter angle and one for total energy, using only those events with exactly one plastic hit and one calorimeter hit. The one plastic hit was required to be the plastic illuminated by the beam at that position. We imposed an energy threshold of 30 keV for the calorimeter and 6 keV for the plastic elements, but placed no constraints on the total energy. In addition, the Compton scattering angle was computed from the plastic and calorimeter energy deposits, and only events that scattered between 45º and 135º were accepted. In this way we maximized the modulation signal while eliminating incompletely absorbed and background events. Each individual histogram, representing one plastic position, was then scaled according to both its exposure time (relative to an arbitrary value of 1000 s) and the count rate recorded in the corresponding beam monitor file.

The scaled histograms were then added together to create a total azimuthal scatter angle histogram and total energy histogram for each energy and polarimeter module orientation. For the azimuthal histogram a bin size of 30º was used to smooth out the sparse angular coverage of the checkerboard pattern. In this way we simulated, for each orientation and energy, a "uniform" exposure of the module to highly polarized radiation. We emphasize that only the relative normalization of the azimuth bins for each beam energy is accurate, and not the absolute normalization, which is arbitrary.

In order to correct each scaled azimuth histogram for geometric effects, we next created an "unpolarized" histogram by adding the scaled azimuth histograms for Orientation 1 and Orientation 2 together. This was used to correct and renormalize the "polarized" histograms as follows:

$$C_{i,corr}^{or1} = \frac{C_i^{or1}}{\left(C_i^{or1} + C_i^{or2}\right)} \times \frac{\sum_i^N \left(C_i^{or1} + C_i^{or2}\right)}{N}, \tag{4}$$

where $C_i^{or1}$ is the number of counts in the $i^{\text{th}}$ bin of the histogram for Orientation 1 (and similarly for Orientation 2) and $N$ is the number of azimuth bins (twelve, for 30º bins). The corrected histograms were then fitted with Eq. 1 and the modulation factor calculated using Eq. 2.

The fitted azimuthal modulation histograms at 69.5 keV are shown in Fig. 7 and 8 for Orientation 1 and 2, respectively, and at 129.5 keV in Fig. 9 and 10. At both energies we obtain the expected sinusoidal modulation pattern and good fits with Eq. 1. The modulation factor, designated $\mu_{97}$ under the assumption that the beam is 97% polarized, is listed for each histogram in the first two rows of Table 1 (labeled "self-corrected" to indicate that the APS data for each orientation has been corrected for geometric effects with the combination of APS data from both orientations using Eq. 4). We obtain the same modulation factor for both orientations, within the errors: $\mu_{97}$ = 0.46 ± 0.06 at 69.5 keV and $\mu_{97}$ = 0.48 ± 0.03 at 129.5 keV. The uncertainty is larger at 69.5 keV, reflective of the larger deviations from the expected sinusoidal pattern evident in Fig. 7 and 8. We believe that this is likely due to a slight contamination of the data both by low-energy room background and by the small number of cross talk events that pass through the PSD filter. We were unable to take a background measurement at APS because the X-Y table did not have sufficient range to move EM1 out of the beam. Both background and cross talk would have less of an effect at 129.5 keV because these events would be less likely to pass the Compton scatter angle cut relative to true beam events.

The total-energy histograms for all selected events at 69.5 keV and 129.5 keV are shown in Fig. 11. The relative normalization of the two histograms is arbitrary for the reasons discussed above. The energy resolution (FWHM) is 24.5% at 69.5 keV and 21.2% at 129.5 keV. The GRAPE instrument is thus well suited for studying energy-dependent polarization. We note, however, that the centroids of the full-energy peaks are shifted to high values: we fit centroid values of 76.2 keV and 131.6 keV instead of 69.5 keV and129.5 keV, respectively. We believe this is due to optical cross talk from the calorimeter: even though the PSD circuit requires valid events to show a fast rise from the plastic elements, the slower signal from the CsI(Na) crystal can contribute a small amount of light to the measured plastic pulse height after the fast coincidence but before the shaped pulse is held for analog-to-digital conversion. For future versions of GRAPE steps will be taken both to minimize and correct for this effect (see Sec. 6).

The same data cuts and analysis procedure described above were applied to the unpolarized data taken using laboratory $^{241}$Am and $^{57}$Co sources (Sec. 3). We thus obtained two independent unpolarized histograms, albeit at slightly different energies than the beam data. The laboratory histograms were used to correct the polarized beam data as before, with the elements of the new unpolarized histogram $C_i^{unpol}$ substituted for the combination $C_i^{or1} + C_i^{or2}$ in Eq. 4. The modulation factors obtained in this way for each orientation are given in the second two rows of Table 1 (labeled "lab-corrected"). Although consistent with the values obtained using Eq. 4 directly, the lab-corrected values vary more widely between the two orientations and have much larger errors, especially at 69.5 keV (e.g., $\mu_{97} = 0.42 \pm 0.16$ for Orientation 1). We take this to mean that, when using unpolarized data to correct for geometric effects, it is important to use data with very good statistics taken at the same energy due to variations in the channel thresholds, gains, and offsets, as discussed further in Sec. 6.

**5. Monte Carlo Simulations**

We have performed detailed Monte Carlo simulations of the APS beam calibration in order to validate our simulation tools and to demonstrate that we fully understand the behavior of the GRAPE instrument. The simulations were performed using the MGGPOD simulation package [34]. MGGPOD is a combination of software tools based on GEANT v3.21; its primary purpose is to simulate background in gamma-ray astronomy instruments, both prompt and delayed, caused by cosmic rays and other particles interacting in detector and spacecraft materials. Although this functionality is beyond what is needed for our present purposes, we plan to use MGGPOD to model the background measured during the GRAPE balloon flight and to predict background in future balloon- and space-based instruments. We therefore use MGGPOD for all our instrument simulation work. We have participated in expanding the functionality of MGGPOD [35], including adding the ability to simulate polarized Compton scattering via the GEANT Low Energy Polarized Scattering (GLEPS) package [36]. As mentioned in Sec. 2.2, MGGPOD simulations using GLEPS have previously been shown to accurately reproduce the laboratory calibration results of the SM2 and SM3 prototypes [29, 30].

We first created a mass model of the GRAPE EM1 detector, including the CsI(Na) and plastic crystals, the VM2000™ wrapping around each crystal, the aluminum CsI(Na) housing, the aluminum module cover, the H8500 MAPMT, and the Delrin™ plastic frame holding the MAPMT. We then simulated $10^5$ photons in a 1 mm diameter beam incident on the center of the same plastic elements illuminated at APS. The polarization fraction was set at 97%. The simulation was repeated for each plastic element with the polarization vector rotated 90º, as in the beam experiment. The individual energy deposits in each active detector element were summed and then randomly altered according to the energy broadening measured in that channel during the EM1 calibration (Sec. 2.3). The measured thresholds for each channel were then applied to determine which detectors elements registered a hit. Finally, the same data cuts were applied as for the APS data (Sec. 4), and azimuthal scatter angle histograms were produced using Eq. 4.

The simulated modulation histogram for 69.5 keV, Orientation 1, is shown in Fig. 12, and for 129.5 keV in Fig. 13. The fitted values of $\mu_{97}$ are given for both orientations in the first two rows of Table 1, next to the values from the beam data. We find $\mu_{97} \approx 0.49 \pm 0.03$ for both energies. This is in excellent agreement with the data at 129.5 keV, and within 5% (and within the measured errors) at 69.5 keV. As noted above, we believe room background and cross talk contribute to the slightly larger measurement errors at 69.5 keV. Unpolarized room background in the data would also slightly reduce the measured modulation factor compared to the simulations, as is observed.

We also simulated the unpolarized lab data at 60 keV and 122 keV, and used it to correct the simulated APS data as described in Sec. 4 for the real data. These results are reported in the last two rows of Table 1, and again show excellent agreement with the measured results. Although the simulated modulation shows less scatter and smaller errors than for the measured data, the simulated "lab-corrected" results are nonetheless slightly noisier than the simulated "self-corrected" results. This further emphasizes the need for the unpolarized correction data to be at the same energy as the corresponding polarized data.

Finally, we repeated the APS simulation using $10^6$ unpolarized photons at precisely 69.5 keV and 129.5 keV in order to test whether high-statistics Monte Carlo simulations of the unpolarized response can be used to correct measured polarized data. The results were quite poor for both energies: for Orientation 2, for example, we find $\mu_{97} = 0.64 \pm 0.19$ at 69.5 keV and $\mu_{97} = 0.37 \pm 0.08$ at 129.5 keV, with significant deviations from the expected sinusoidal pattern at both energies (e.g., Fig. 14).

**6. Conclusions and Future Plans**

We have demonstrated that the GRAPE polarimeter design is capable of achieving high values of $\mu_{100}$, near 0.5, at hard X-ray energies; that GRAPE has sufficient energy resolution, ~20 – 24% (FWHM), to measure polarization as a function of energy; and that our Monte Carlo tools, based on MGGPOD and the GLEPS package, correctly describe the performance of the real instrument. We are therefore confident that we can develop highly-sensitive balloon- and space-based instruments based on the GRAPE polarimeter module, and that we can accurately predict their scientific performance.

There are, however, several important lessons to be drawn from our experience with the APS data and simulations. The first is that the correction of the polarized data for geometric asymmetries depends sensitively on the response of the instrument, and must therefore be performed with great care. Our best results were obtained using the beam data itself to create an unpolarized histogram. Using unpolarized laboratory data taken at slightly different energies gave significantly worse results, especially at low energy, while using Monte Carlo simulations alone gave even worse results. We believe that this is due to insufficient knowledge of the low-energy thresholds and response (gain, offset, and resolution) of the detector elements. The lowest-energy calibration point for the plastic elements was 22 keV, well above the desired threshold, and the energy calibration was based on only two points measured with relatively low statistics due to time constraints. The poor energy resolution of the plastic scintillator means both that calibration data must be taken with high statistics in order to accurately fit the energy response, and that a significant number of events will always be lost below threshold at low energies. Therefore, in order to correctly model the instrument and apply the geometric corrections, the thresholds, energy calibration, and energy broadening must all be accurately measured and included in the simulations. In this case, the differences between the true thresholds and those assumed in the simulations were significant enough to lead to an unrealistic simulated response to unpolarized radiation, and thus poorly reconstructed modulation factors.

A corollary is that the energy response and thresholds should be as uniform as possible amongst all detector elements. The thresholds in the EM1 detector varied between 6 keV and 12 keV due to differences in the light collection, gain, and electronic noise. These variations themselves introduce an artificial asymmetry in the azimuthal response of the polarimeter that varies with energy, becoming most pronounced at energies near threshold. We believe this is the cause of the poor results using 60 keV unpolarized lab data to correct the polarized 69.5 keV beam data: even this small difference in energy produced significantly different azimuthal scatter patterns due to the varying thresholds of the plastic elements. Specifically, 60 keV photons Compton scattering between 45º and 135º will deposit roughly between 2 keV and 10 keV in the plastic elements; since this energy range spans the range of plastic thresholds, there will be a large variation in the detection efficiency of the different elements, especially when the poor energy resolution of the plastic is taken into account. In principle one could impose a uniform threshold, either in hardware or in software, that lies above that of all detector elements; however, since the nominal energy range of GRAPE extends down to 50 keV, it is highly desirable to have the lowest achievable threshold in as many plastic elements as possible in order to maximize the low-energy detection efficiency. We note that most astrophysical objects are brightest at low energies, and so maintaining a low overall threshold is critical for high-sensitivity polarization measurements.

These considerations lead us to the conclusion that the best way to correct for geometrical asymmetries is the second option identified in Sec. 2.1: to physically rotate the polarimeter around its azimuthal axis. In this scenario, each recorded event is tagged with the instantaneous azimuth angle of the instrument so that the scattering angle may be transformed back into the laboratory frame. In this way both geometric asymmetries and variations in gains and thresholds are averaged out over the full 360º azimuth range in real time, leaving only the true polarization signal in the laboratory frame superimposed on the detector background. No subsequent correction of the azimuth histogram is required; only the DC background level need be subtracted off. Rotating EM1 by 90º at APS represented a first step in this direction, since it allowed us to remove the effects of both the 90º instrument symmetry and the varying gains and thresholds without measuring them precisely. More complete sampling over a full 360º would have further removed the effects of non-symmetric cross talk events. The advantages of rotating a Compton polarimeter have been noted by other authors as well (e.g., [12, 31, 33]).

We note that calibration measurements and simulations are still necessary in order to measure the absolute polarization fraction Π, since the magnitude of $\mu_{100}$ will depend on the instrument-specific detector configuration. This is especially true for polarized sources that lie significantly off-axis relative to the rotation. In addition, if the intensity of the source varies on a timescale shorter than the rotation period, as is likely for some GRBs, then this must also be accounted for in the analysis.

Another lesson learned from the APS beam campaign is that, for future iterations of the GRAPE instrument, great care must be taken to minimize, and correct for, cross talk between calorimeter and plastic elements. Cross talk at some level is most likely unavoidable while using the H8500 MAPMT due to its 2 mm-thick entrance window, which allows light from each crystal to reach the neighboring anode, and internal capacitance (an inherent electronic cross talk level of 3% is quoted in the H8500 technical specifications sheet). The use of a calorimeter material with lower light output, such as BGO, would reduce the magnitude of the spurious signals in the plastic channels, at the cost of degraded energy resolution. Completely removing the cross talk signal from the plastic channels could perhaps be accomplished by a more complex readout system, such as digitizing the individual pulse shape for each event to allow for fitting and subtracting the slower CsI(Na) pulse. Such schemes are likely to require a prohibitive amount of power, however. A more practical approach may simply be to continue using PSD to prevent most cross talk events from triggering the system, and to perform the energy calibration of each plastic element in polarimeter mode rather than in singles mode. Using the known input energy and the measured calorimeter energy, the true energy deposit in the plastic element may then be calculated and related to the measured plastic pulse height.

It is also clear that great care must be taken to achieve the lowest possible threshold for each plastic element, to make these thresholds as uniform as possible, and to accurately measure the energy calibration all the way down to the threshold. In order to detect a 50 keV photon Compton scattering at 45º, for example, a threshold of 1.4 keV is necessary. We will continue to study the optimum reflective materials, surface treatments, and optical coupling methods in order to maximize the light collection from the plastic elements. More advanced photocathode materials, such as the "Super Bialkali" photocathode recently announced by Hamamatsu, offer the promise of increased quantum efficiency and thus lower thresholds as well. We have also identified an improvement for future versions of the electronics that will eliminate the noise inherent in conventional CFD or pulse shape discriminators and should permit a lower threshold. For our future hardware development we will attempt to extend the calibration of each plastic element down to the 5.9 keV line of $^{55}$Fe in order to guarantee a low, well-measured threshold. This will in turn assure good detection efficiency for GRAPE down to at least 50 keV.

Based on Monte Carlo simulations, the next generation of the GRAPE polarimeter module will employ 28 calorimeter elements placed around the outside of the H8500 MAPMT, as depicted in Fig. 1. This will roughly double the detection efficiency compared to SM3, due to the much larger probability of scattered photons hitting a calorimeter, and also increase the modulation factor by ~15%, due to the larger average distance between plastic and calorimeter hits and the smaller angular size of the calorimeter elements. We will continue to improve the light collection and energy calibration of the plastic elements and to study methods for reducing, and correcting for, cross talk, as discussed above. All future designs of the GRAPE instrument will incorporate rotation through at least 180º to correct for systematic geometric effects. Finally, we will consider different methods of using GRAPE in an imaging polarimeter system.

Based on the results of the APS beam campaign and our Monte Carlo simulations we are confident that GRAPE will be a sensitive hard X-ray polarimeter for astronomical observations of GRBs and solar flares. We have proposed a rotating array of 36 next-generation GRAPE modules as a long-duration scientific balloon payload, and will continue to pursue further

balloon- and space-flight opportunities.

**Acknowledgements**

We would like to thank P. Vachon and D. Rhines for their support of the GRAPE readout electronics, M. Chutter for the data acquisition software, and C. Seymour for assistance with packing. We would also like to thank D. Robinson of the MUCAT sector at APS for his invaluable assistance in making the beam campaign a success. Use of the Advanced Photon Source at Argonne National Laboratory was supported by the U.S. Department of Energy, Office of Science, Office of Basic Energy Sciences, under Contract No. DE-AC02-06CH11357. The Midwest Universities Collaborative Access Team (MUCAT) sector at the APS is supported by the U.S. Department of Energy, Office of Science, Office of Basic Energy Sciences, through the Ames Laboratory under Contract No. DE-ACD2-07CH11358. This work was supported by NASA grants NNG04GB83G and NNG04WC16G.

**References**

[1] F. Lei, A.J. Dean, G.L. Hills, Sp. Sci. Rev. 82 (1997) 309.

[2] T. Piran, Rev. Mod. Phys. 76 (2005) 1143.

[3] P. Meszaros, Rep. Prog. Phys. 69 (2006) 2259.

[4] E. Waxman, Nature 423 (2003) 388.

[5] R. P. Lin, H. S. Hudson, Sol. Phys. 50 (1976) 153.

[6] G. Chanan, et al., Sol Phys. 118 (1988) 309.

[7] T. Bai, R. Ramaty, Astrophys. Jour. 219 (1978) 705.

[8] J. Leach, V. Petrosian, Astrophys. Jour. 269 (1983) 715.

[9] J. R. P. Angel, et al., Phys. Rev. Lett. 22 (1969) 861.

[10] I. P. Tindo, et al., Sol. Phys. 14 (1970) 204.

[11] M. C. Weisskopf, et al., Astrophys. Jour. 220 (1978) L117.

[12] W. Coburn, S. E. Boggs, Nature 423 (2003) 415.

[13] D. R. Willis, et al., Astron. Astrophys. 439 (2005) 245.

[14] E. Kalemci, et al., Astrophys. Jour. Supp. Ser. 169 (2007) 75.

[15] S. McGlynn., et al., Astron. Astrophys. 466 (2007) 895.

[16] M. L. McConnell, et al., Bull. Amer. Astron. Soc. 34 (2003) 850.
[17] A. V. Bogomolov, et al., Sol. Sys. Res. 37 (2003) 112.
[18] S. E. Boggs, W. Coburn, E. Kalemci, Astrophys. Jour. 638 (2006) 1129.
[19] E. Suarez-Garcia, et al., Sol. Phys. 239 (2006) 149.
[20] I. A. Zhitnik, et al., Sol. Sys. Res. 40 (2006) 93.
[21] R. E. Rutledge, D. B. Fox, Mon. Not. Roy. Astron. Soc. 350 (2004) 1272.
[22] C. Wigger, et al., Astrophys. Jour. 613 (2004) 1088.
[23] M. L. McConnell, et al., AIP Conf. Ser. 428 (1998) 889.
[24] M. L. McConnell, et al., IEEE Trans. Nucl. Sci. 45 (1998) 910.
[25] M. L. McConnell, et al., Proc. SPIE 3764 (1999) 70.
[26] M. L. McConnell, et al., IEEE Trans. Nucl. Sci. 46 (1999) 890.
[27] M. L. McConnell, et al., Proc. SPIE 4851 (2003) 1382.
[28] M. L. McConnell, et al., Proc. SPIE 5165 (2004) 334.
[29] J. Legere, et al., Proc. SPIE 5898 (2005) 413.
[30] P. F. Bloser, et al., Ch. Jour. Astron. Astrophys. 6 (2006) 393.
[31] Y. Kanai, et al., Nucl. Instr. and Meth. A 570 (2007) 61.
[32] N. Produit, et al., Nucl. Instr. and Meth. A 550 (2005) 616.
[33] Y. Kishimoto, et al., IEEE Trans. Nucl. Sci. 54 (2007) 561.
[34] G. Weidenspointner, et al., Astrophys. Jour. Supp. Ser. 156 (2005) 69.
[35] C. B. Wunderer, et al., New Astron. Rev. 50 (2006) 608.
[36] M. L. McConnell, R. M. Kippen, P. F. Bloser, in preparation.

**Table 1: Summary of GRAPE modulation factors for 97% polarized radiation ($\mu_{97}$): APS data vs. simulations.**

| | 69.5 keV | | 129.5 keV | |
|---|---|---|---|---|
| | APS Data | Simulation | APS Data | Simulation |
| Self-corrected, Orientation 1 | 0.46 ± 0.05 | 0.49 ± 0.03 | 0.48 ± 0.03 | 0.49 ± 0.02 |
| Self-corrected, Orientation 2 | 0.46 ± 0.06 | 0.48 ± 0.02 | 0.48 ± 0.02 | 0.49 ± 0.02 |
| Lab-corrected, Orientation 1 | 0.54 ± 0.17 | 0.50 ± 0.05 | 0.52 ± 0.05 | 0.50 ± 0.02 |
| Lab-corrected, Orientation 2 | 0.42 ± 0.16 | 0.48 ± 0.05 | 0.46 ± 0.05 | 0.48 ± 0.02 |

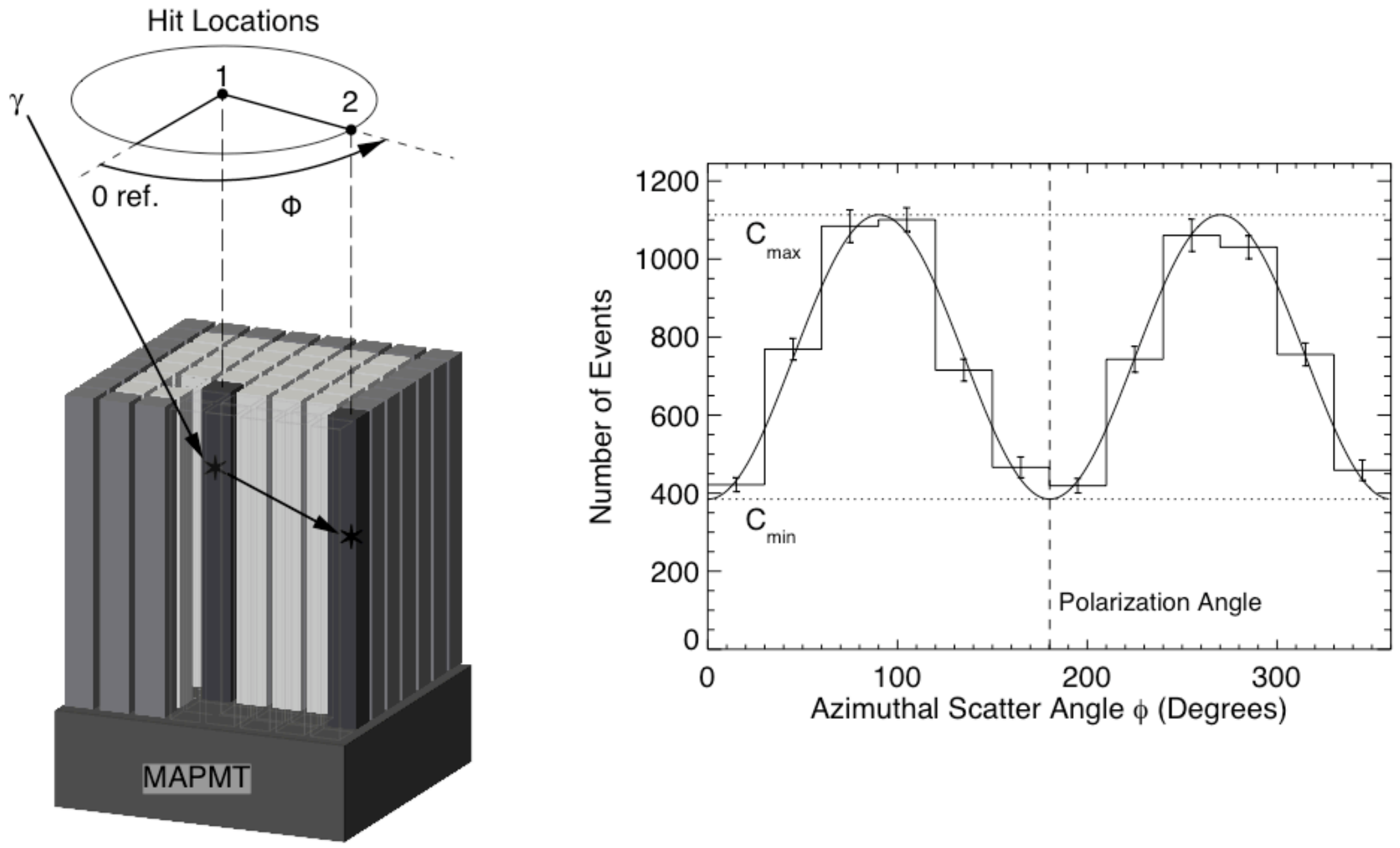

Fig. 1. Principle of operation of the GRAPE polarimeter. Left: An incident hard X-ray photon ($\gamma$) Compton scatters in a plastic (light grey) scintillator element and is absorbed in a calorimeter (dark grey) element. The relative positions give the azimuthal scatter angle ($\phi$). Right: The histogram of $\phi$ for all events reveals the modulation pattern, which is fit with Eq. 1 to find the modulation factor (Eq. 2) and polarization angle.

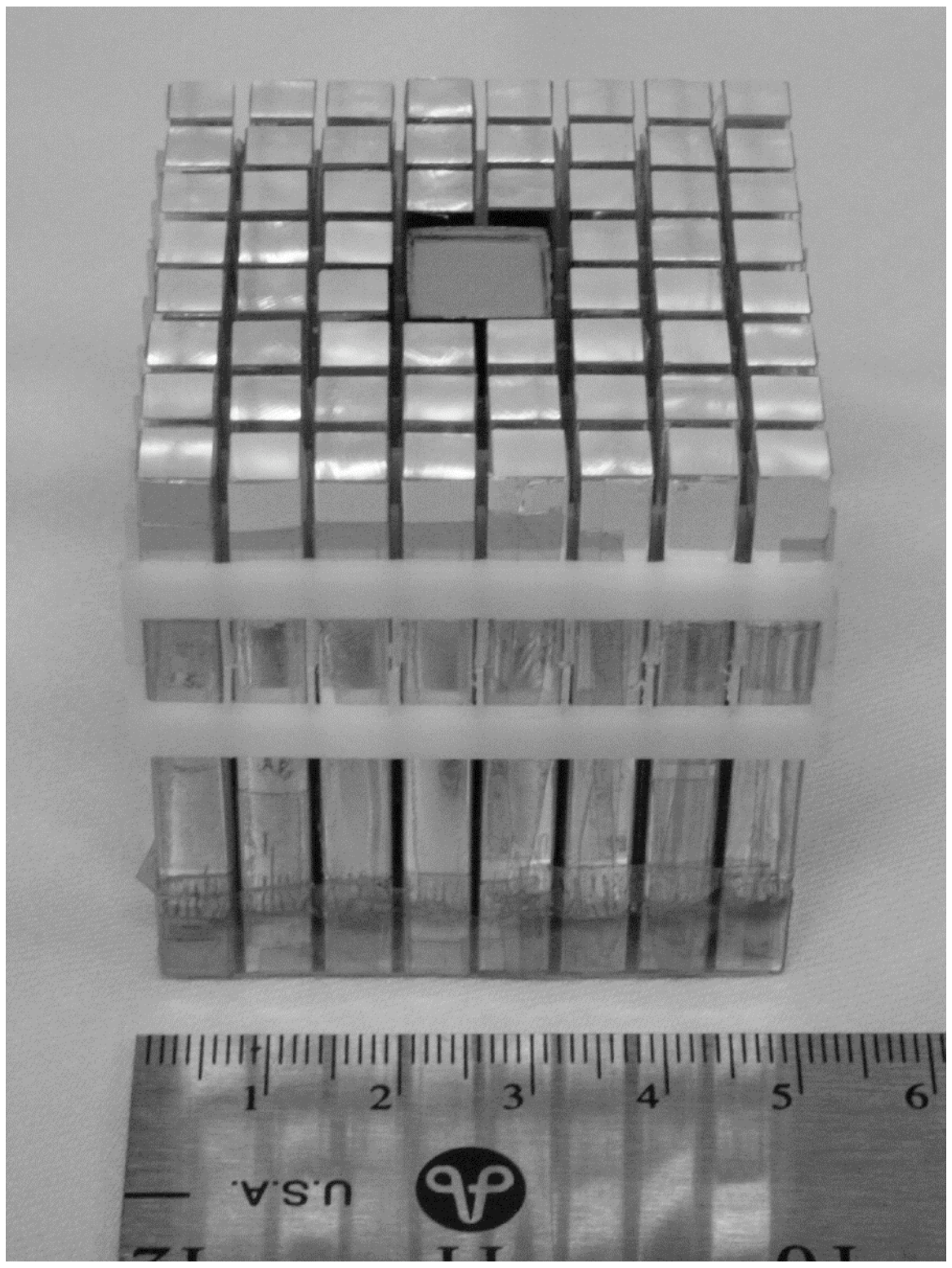

Fig. 2.  The third Science Model (SM3) prototype GRAPE polarimeter module.  The central 10 mm × 10 mm × 50 mm CsI(Na) calorimeter is surrounded by 60 plastic scattering elements, each 5 mm × 5 mm × 50 mm.

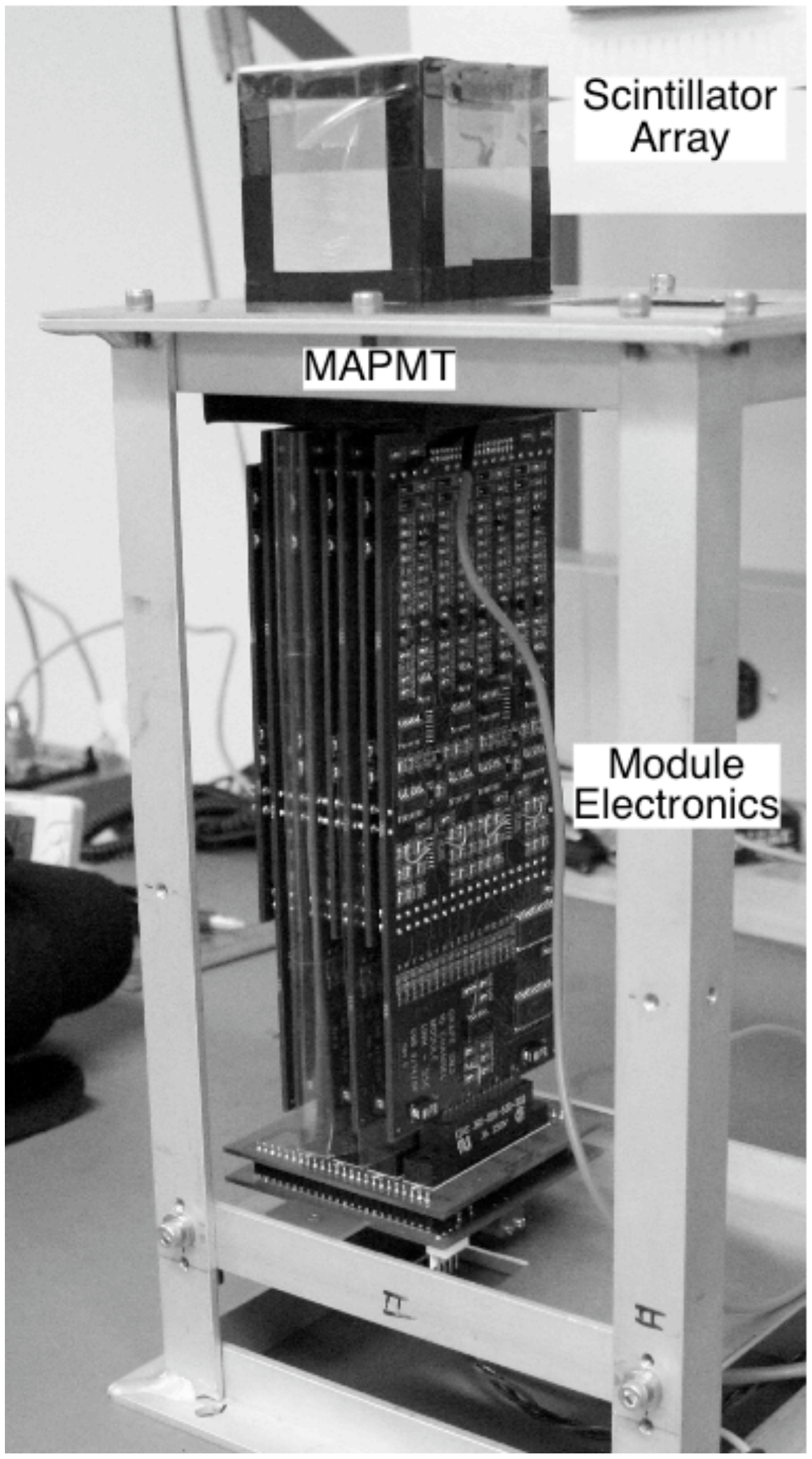

Fig. 3. The GRAPE Engineering Model (EM1). The SM3 polarimeter module is coupled to an MAPMT and read out by custom electronics and a dedicated PC/104 computer (not shown).

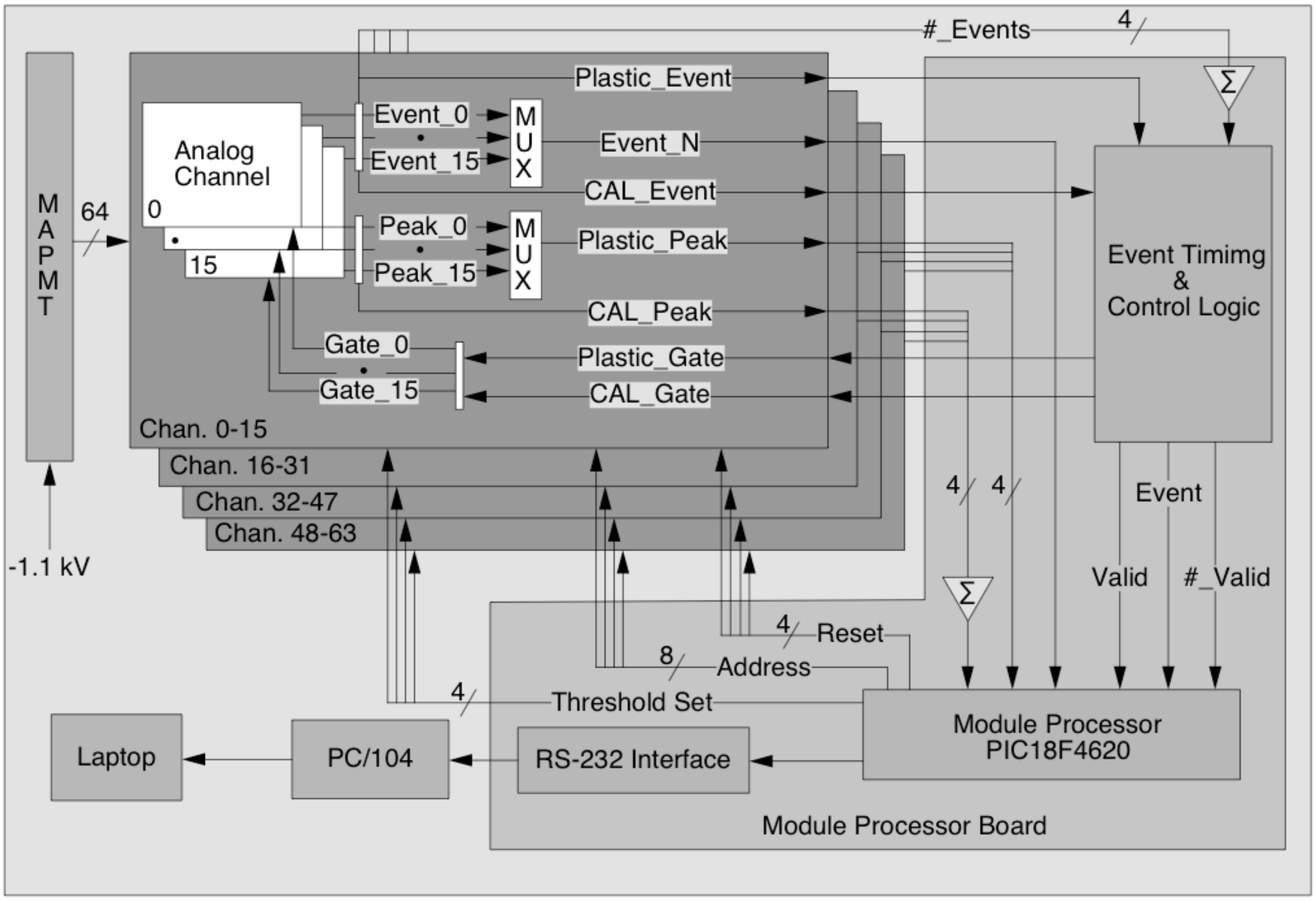

Fig. 4. Block diagram of the EM1 readout electronics. See text for details.

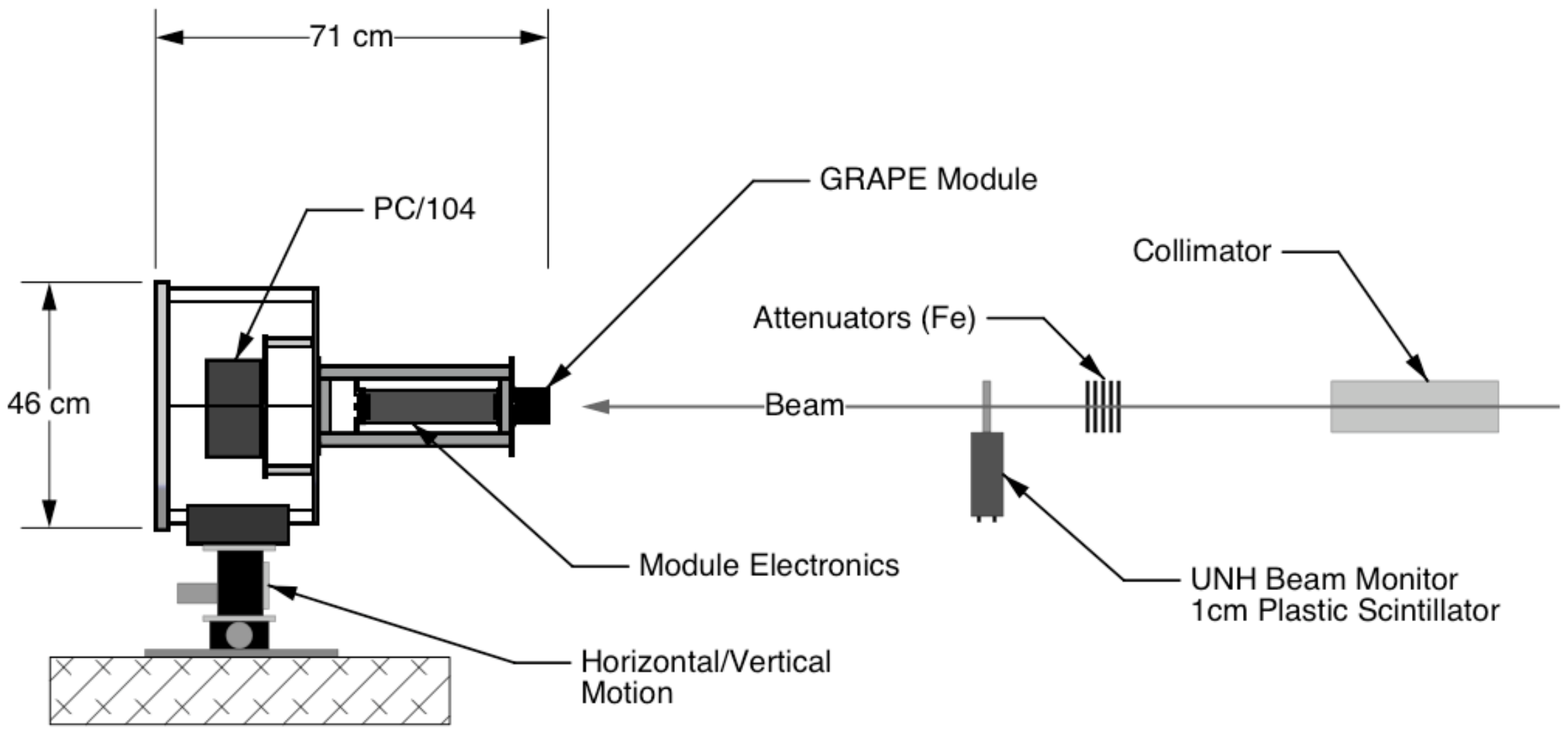

Fig. 5. Schematic view of the experimental setup at APS. EM1 was mounted on an X-Y table to scan plastic elements through the beam one at a time. The beam flux was adjusted by varying the number of attenuators, and monitored by a 1 cm-thick plastic scintillator paddle.

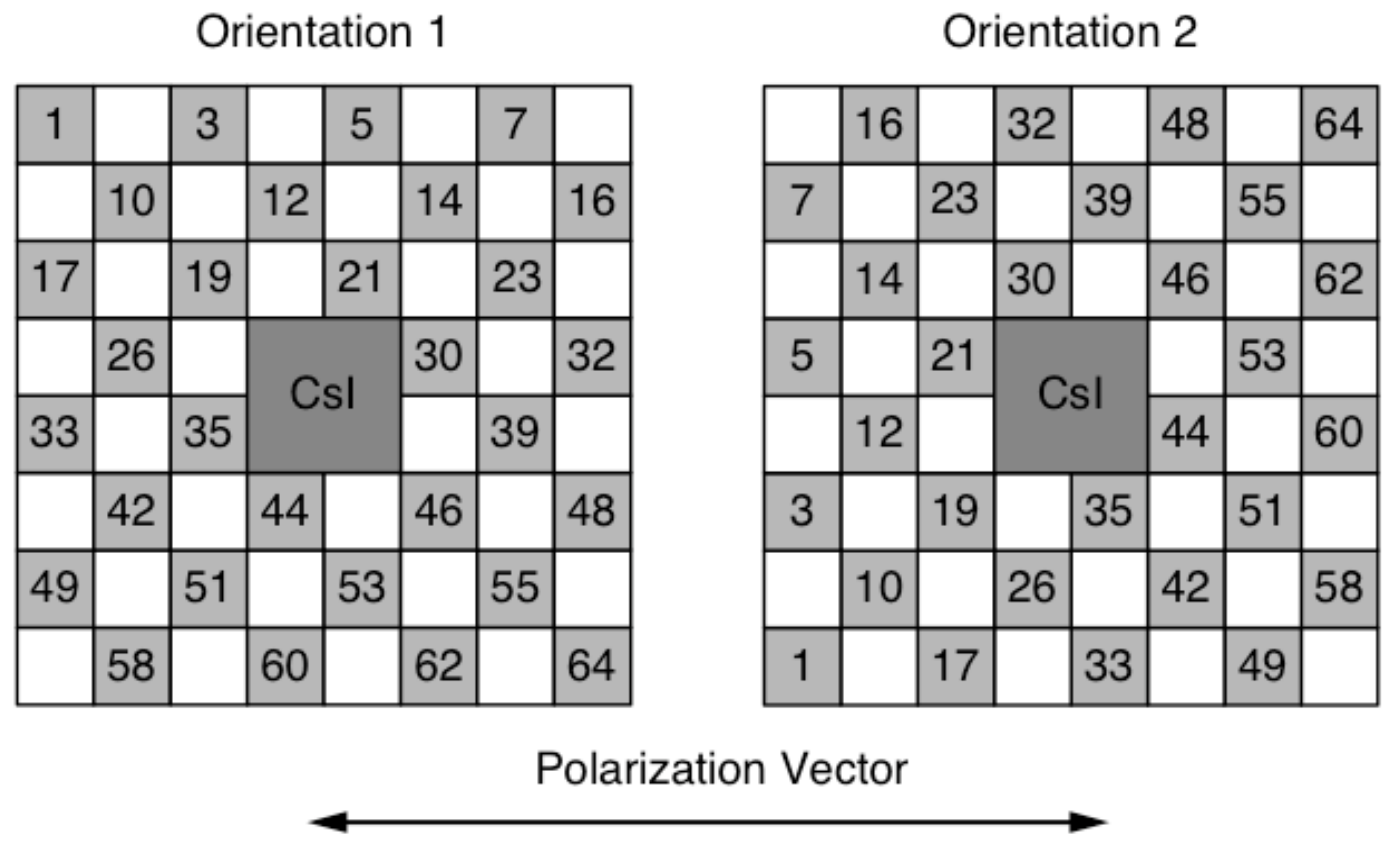

Fig. 6. Exposure pattern for the GRAPE polarimeter module at APS. The numbered plastic elements were exposed to the polarized beam (horizontal polarization vector) until ~10,000 events in coincidence with the CsI(Na) were recorded for each. The module was exposed in two orientations, rotated 90º from each other.

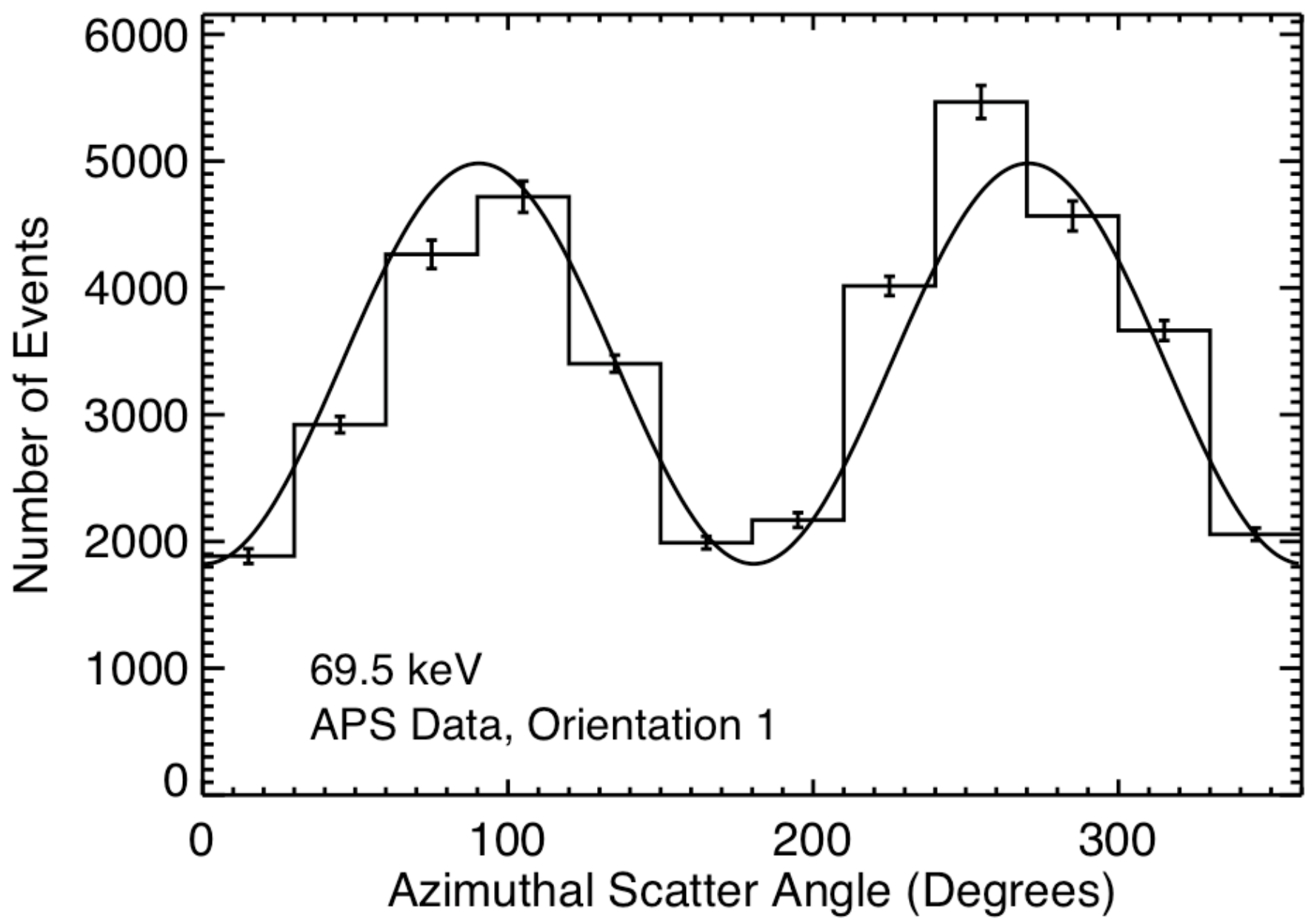

Fig. 7. Azimuthal scatter angle histogram for the GRAPE polarimeter module in Orientation 1 obtained with APS beam data at 69.5 keV using Eq. 4. The fitted modulation factor $\mu_{97}$ is 0.46 ± 0.05 and the polarization angle $\phi_0$ (relative to the instrument) is 0.5º ± 2.5º.

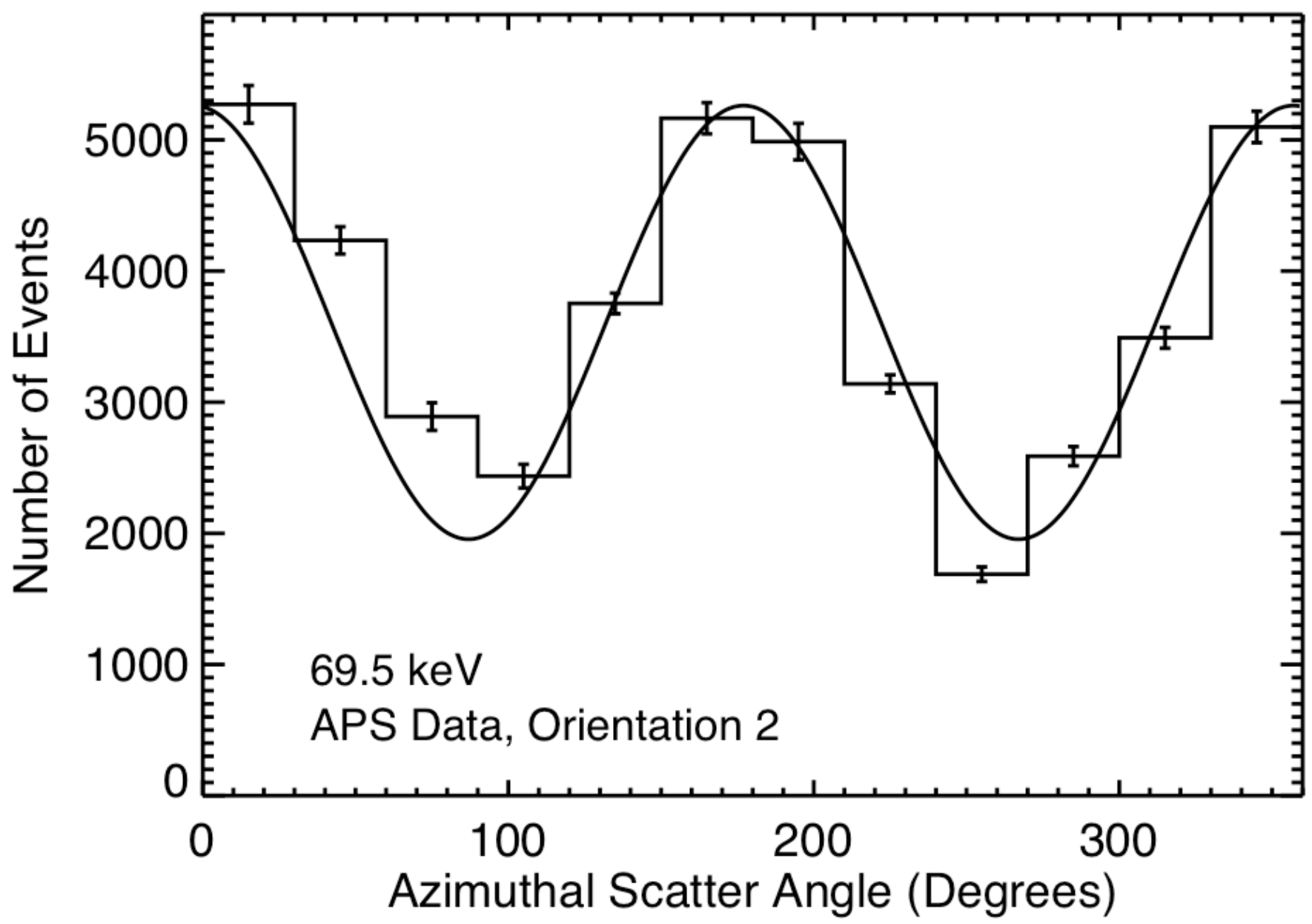

Fig. 8. Azimuthal scatter angle histogram for the APS beam at 69.5 keV in Orientation 2. We find $\mu_{97} = 0.46 \pm 0.06$ and $\phi_0 = 87.0^{\circ} \pm 3.0^{\circ}$.

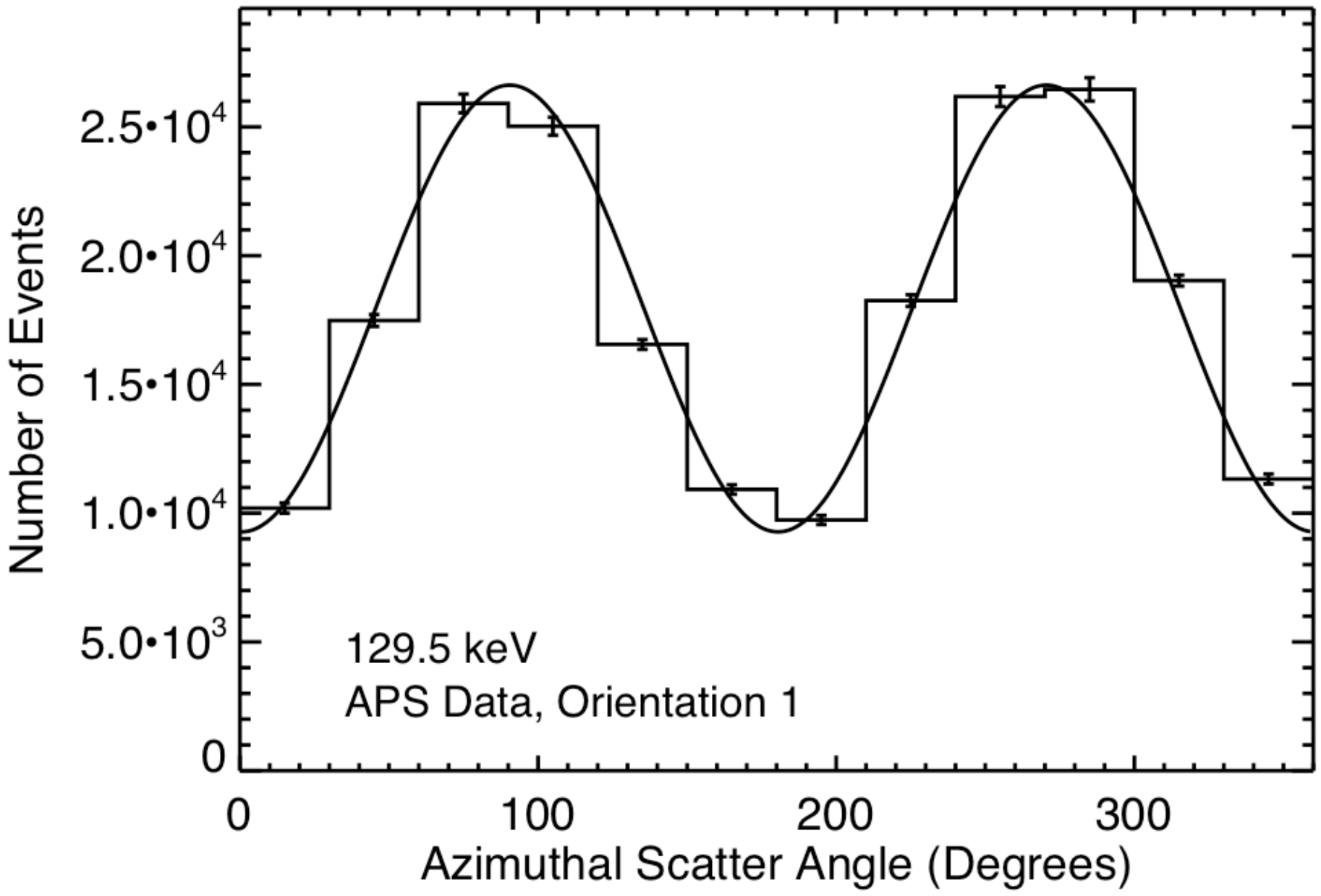

Fig. 9. Azimuthal scatter angle histogram for the APS beam at 129.5 keV in Orientation 1. We find $\mu_{97} = 0.48 \pm 0.03$ and $\phi_0 = 0.5^o \pm 1.2^o$.

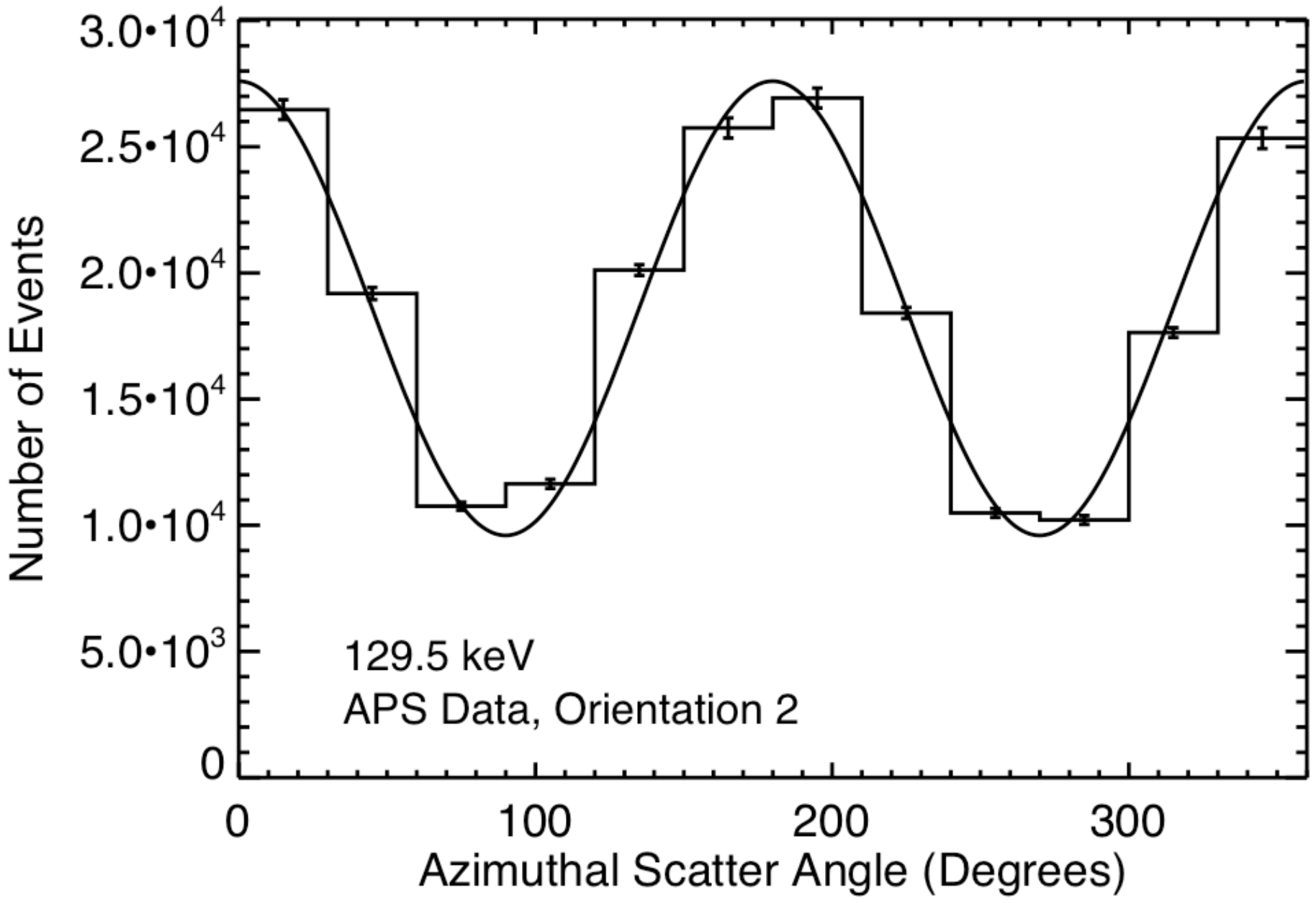

Fig. 10. Azimuthal scatter angle histogram for the APS beam at 129.5 keV in Orientation 2. We find $\mu_{97} = 0.48 \pm 0.02$ and $\phi_0 = 89.9^o \pm 1.1^o$.

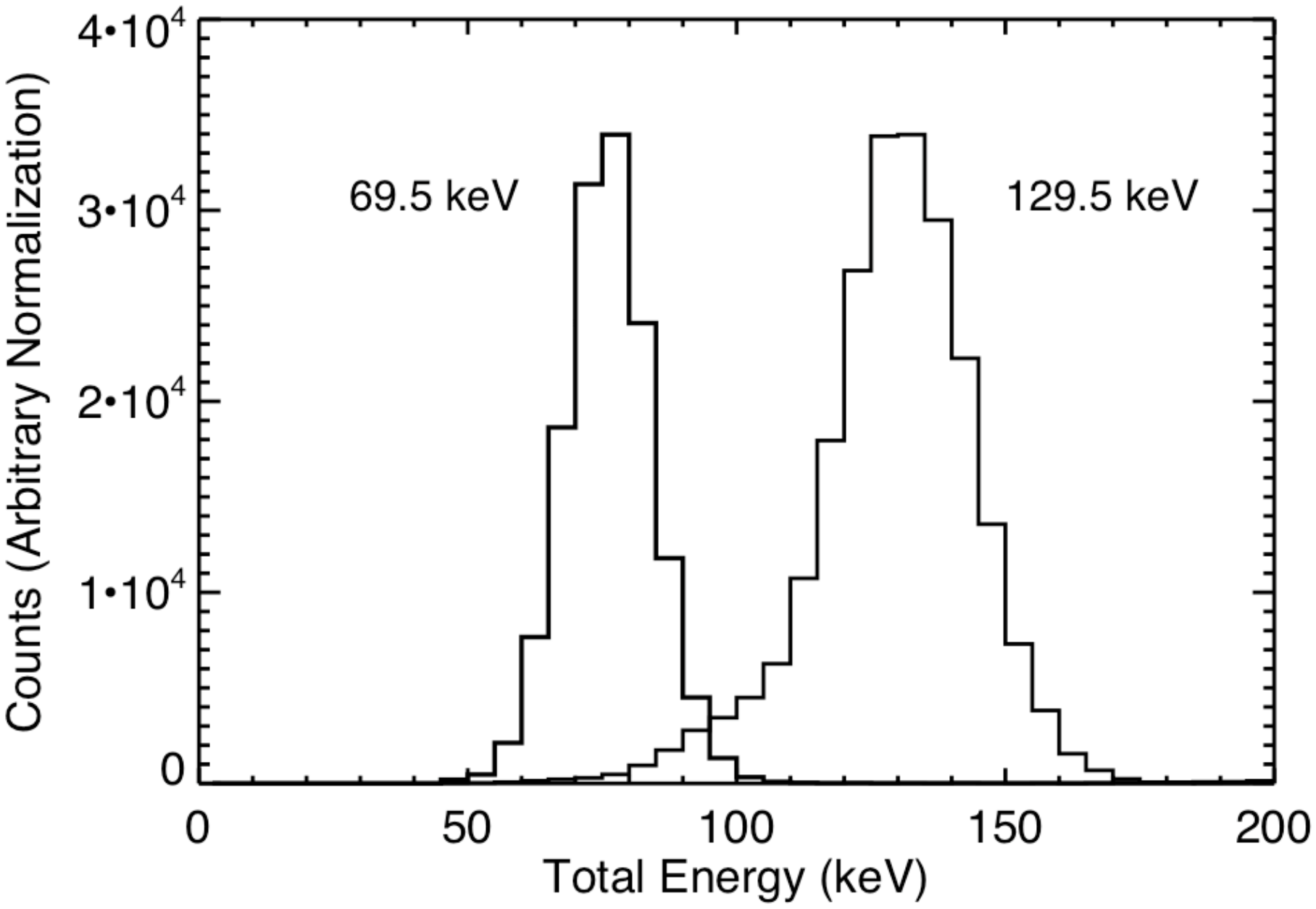

Fig. 11. Histograms of the total energy recorded by the GRAPE module at APS for both the 69.5 keV and 129.5 keV runs. The relative normalization of the histograms is arbitrary. The energy resolution (FWHM) is 24.5% at 69.5 keV and 21.2% at 129.5 keV. The centroids of the full-energy peaks are shifted to slightly high values (76.2 keV and 131.6 keV, respectively) due to cross talk effects (see text).

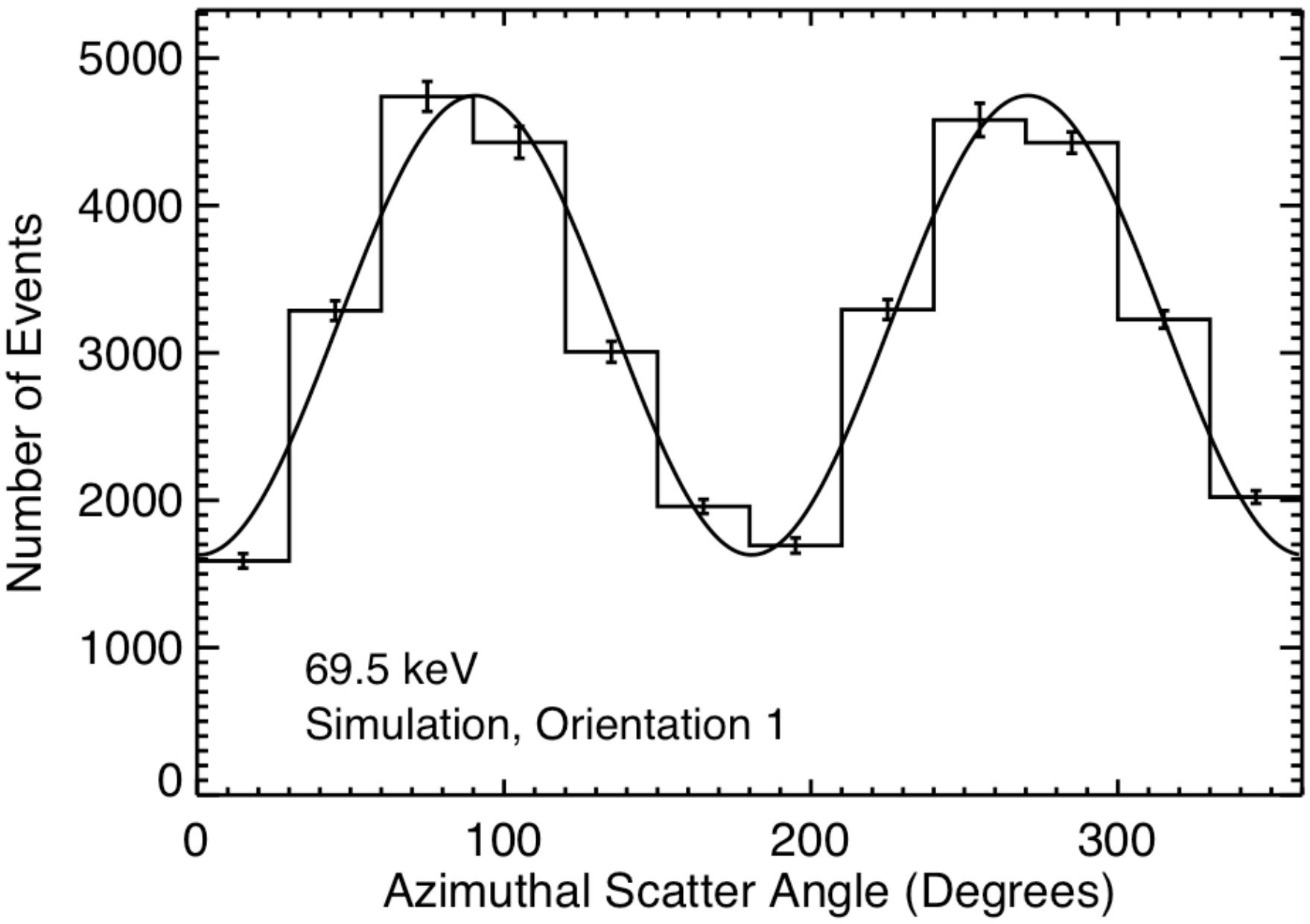

Fig. 12. Simulated azimuthal scatter angle histogram for the 69.5 keV APS beam run (Orientation 1). The simulated data were analyzed in the same manner as the beam data (see text). We find $\mu_{97} = 0.49 \pm 0.03$ and $\phi_0 = 0.7^{\circ} \pm 1.4^{\circ}$.

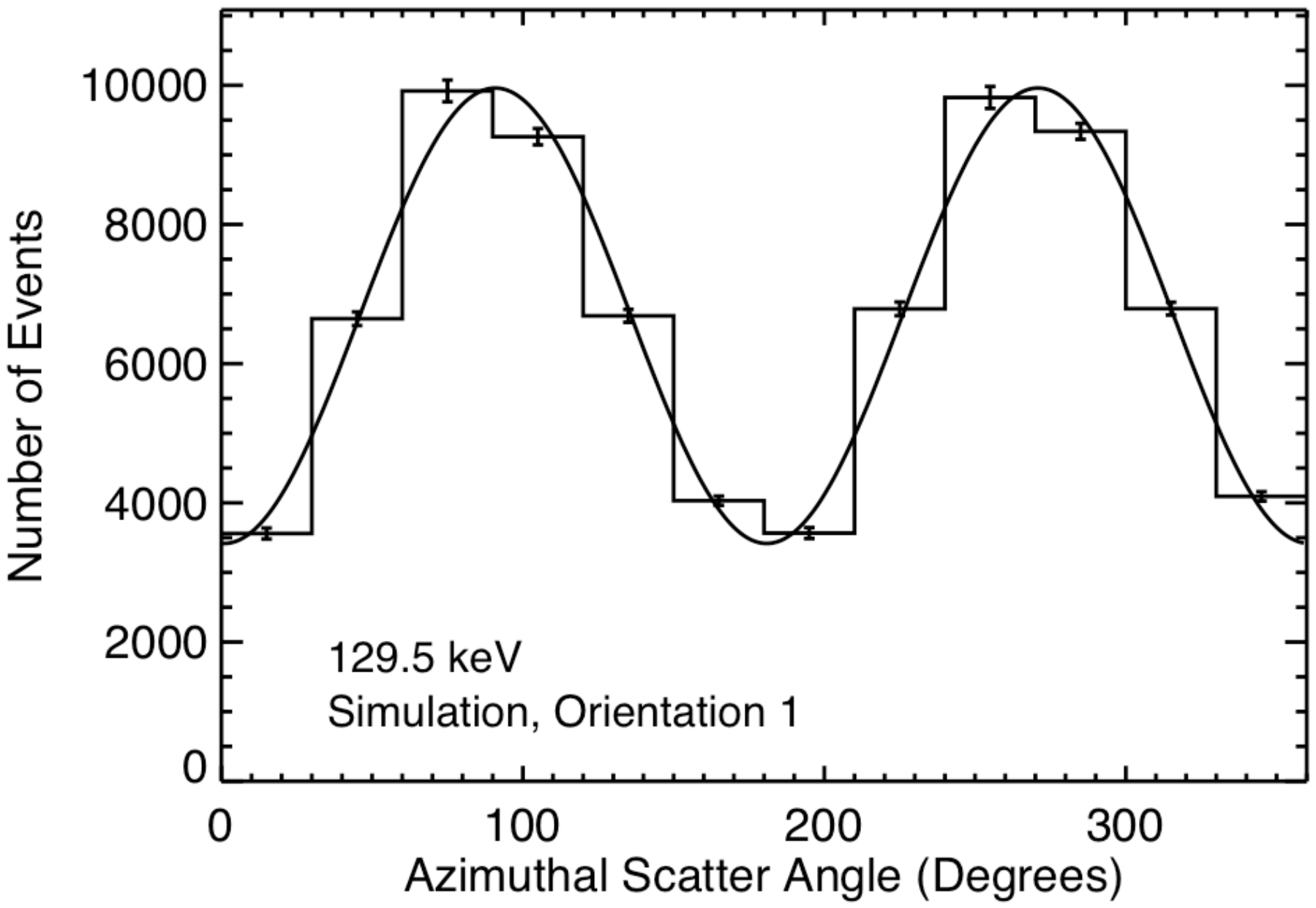

Fig. 13. Simulated azimuthal scatter angle histogram for the 129.5 keV APS beam run (Orientation 1). We find $\mu_{97} = 0.49 \pm 0.02$ and $\phi_0 = 0.9^{o} \pm 0.9^{o}$.

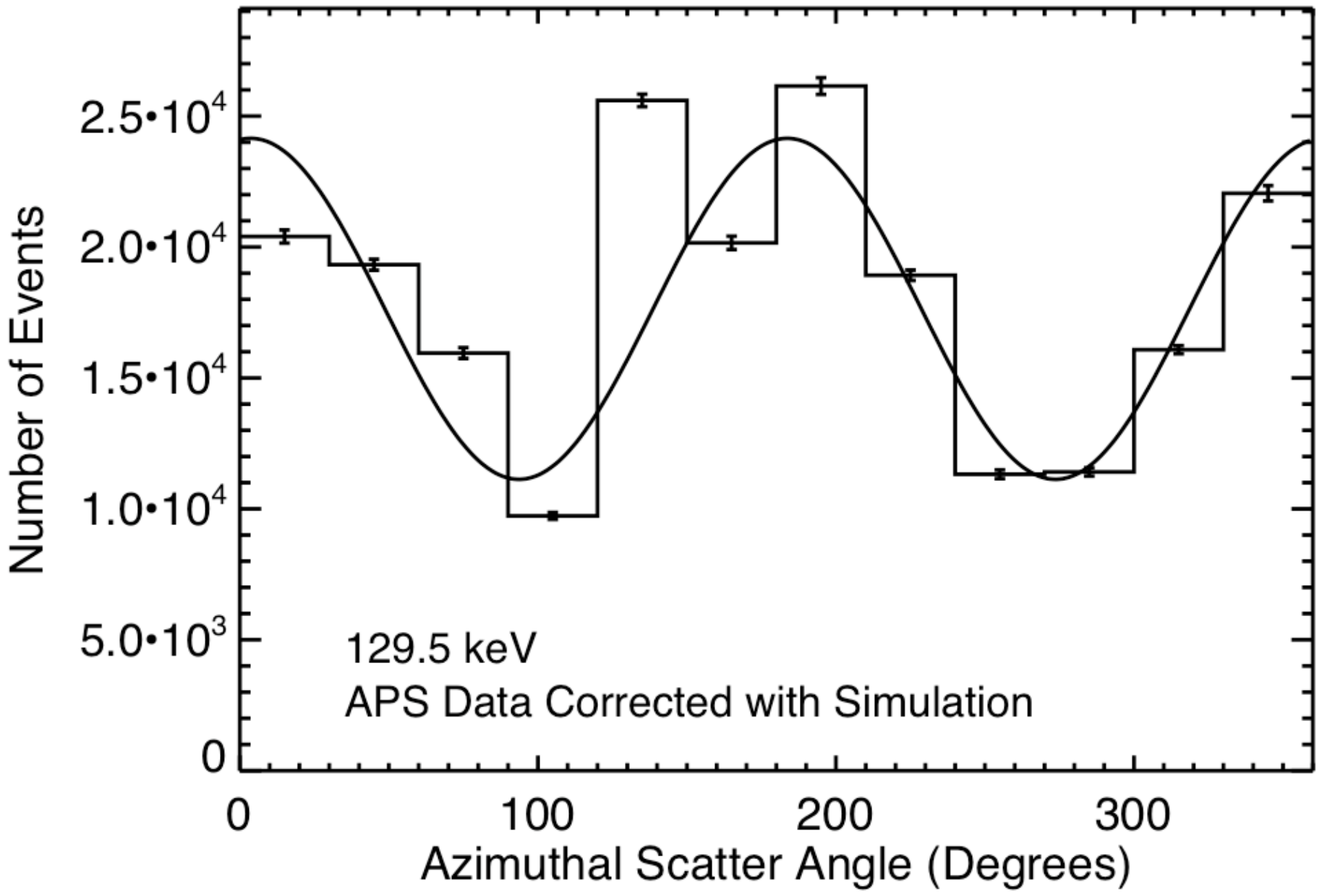

Fig. 14. Azimuthal scatter angle histogram for the APS beam at 129.5 keV in Orientation 2, corrected for geometric effects using simulated unpolarized data. We find $\mu_{97}$ = 0.37 ± 0.08 and $\phi_0$ = 93.7º ± 5.6º. This results are poor compared to those obtained by correcting using the combination of APS data in both orientations (Fig. 10).